\documentclass[twocolumn]{aastex631}
\usepackage{apjfonts, natbib}

\newcommand{\mbh}{\ensuremath{M_\mathrm{BH}}}

\usepackage{epsfig,graphics,subfigure,psfrag,amsmath,amssymb,amscd}

\begin{document}

\title{AT~2023lli: A Tidal Disruption Event with Prominent Optical Early Bump and Delayed Episodic X-ray Emission}

\author[0000-0001-7689-6382]{Shifeng Huang}
\affiliation{CAS Key Laboratory for Research in Galaxies and Cosmology, Department of Astronomy, University of Science and Technology of China, Hefei, 230026, China; sfhuang999@ustc.edu.cn, jnac@ustc.edu.cn,twang@ustc.edu.cn}
\affiliation{School of Astronomy and Space Sciences,
University of Science and Technology of China, Hefei, 230026, China}

\author[0000-0002-7152-3621]{Ning Jiang}
\affiliation{CAS Key Laboratory for Research in Galaxies and Cosmology, Department of Astronomy, University of Science and Technology of China, Hefei, 230026, China; sfhuang999@ustc.edu.cn, jnac@ustc.edu.cn,twang@ustc.edu.cn}
\affiliation{School of Astronomy and Space Sciences,
University of Science and Technology of China, Hefei, 230026, China}

\author[0000-0003-3824-9496]{Jiazheng Zhu}
\affiliation{CAS Key Laboratory for Research in Galaxies and Cosmology, Department of Astronomy, University of Science and Technology of China, Hefei, 230026, China; sfhuang999@ustc.edu.cn, jnac@ustc.edu.cn,twang@ustc.edu.cn}
\affiliation{School of Astronomy and Space Sciences,
University of Science and Technology of China, Hefei, 230026, China}

\author[0000-0003-4225-5442]{Yibo~Wang}
\affiliation{CAS Key Laboratory for Research in Galaxies and Cosmology, Department of Astronomy, University of Science and Technology of China, Hefei, 230026, China; sfhuang999@ustc.edu.cn, jnac@ustc.edu.cn,twang@ustc.edu.cn}
\affiliation{School of Astronomy and Space Sciences,
University of Science and Technology of China, Hefei, 230026, China}

\author[0000-0002-1517-6792]{Tinggui Wang}
\affiliation{CAS Key Laboratory for Research in Galaxies and Cosmology, Department of Astronomy, University of Science and Technology of China, Hefei, 230026, China; sfhuang999@ustc.edu.cn, jnac@ustc.edu.cn,twang@ustc.edu.cn}
\affiliation{School of Astronomy and Space Sciences,
University of Science and Technology of China, Hefei, 230026, China}
\affiliation{Institute of Deep Space Sciences, Deep Space Exploration Laboratory, Hefei 230026, China}

\author[0000-0001-7867-9912]{Shan-Qin Wang}
\affiliation{Guangxi Key Laboratory for Relativistic Astrophysics, School of Physical Science and Technology, Guangxi University, Nanning 530004, People's Republic of
China}

\author{Wen-Pei Gan}
\affiliation{Nanjing Hopes Technology Co., Ltd.  Pukou District
Nanjing, 210000, China}

\author[0000-0002-7044-733X]{En-Wei Liang}
\affiliation{Guangxi Key Laboratory for Relativistic Astrophysics, School of Physical Science and Technology, Guangxi University, Nanning 530004, People's Republic of China}

\author[0000-0003-3658-6026]{Yu-Jing Qin}
\affiliation{Division of Physics, Mathematics and Astronomy, California Institute of Technology, 1200 E California Blvd., Pasadena, CA 91125, USA}

\author[0000-0003-4959-1625]{Zheyu Lin}
\affiliation{CAS Key Laboratory for Research in Galaxies and Cosmology, Department of Astronomy, University of Science and Technology of China, Hefei, 230026, China; sfhuang999@ustc.edu.cn, jnac@ustc.edu.cn,twang@ustc.edu.cn}
\affiliation{School of Astronomy and Space Sciences,
University of Science and Technology of China, Hefei, 230026, China}

\author[0009-0006-4524-4414]{Lin-Na Xu}
\affiliation{Guangxi Key Laboratory for Relativistic Astrophysics, School of Physical Science and Technology, Guangxi University, Nanning 530004, People's Republic of China}

\author[0000-0003-4721-6477]{Min-Xuan~Cai}
\affiliation{CAS Key Laboratory for Research in Galaxies and Cosmology, Department of Astronomy, University of Science and Technology of China, Hefei, 230026, China; sfhuang999@ustc.edu.cn, jnac@ustc.edu.cn,twang@ustc.edu.cn}
\affiliation{School of Astronomy and Space Sciences,
University of Science and Technology of China, Hefei, 230026, China}
\author[0000-0002-9092-0593]{Ji-an~Jiang}
\affiliation{CAS Key Laboratory for Research in Galaxies and Cosmology, Department of Astronomy, University of Science and Technology of China, Hefei, 230026, China; sfhuang999@ustc.edu.cn, jnac@ustc.edu.cn,twang@ustc.edu.cn}
\affiliation{School of Astronomy and Space Sciences,
University of Science and Technology of China, Hefei, 230026, China}
\affiliation{National Astronomical Observatory of Japan, National Institutes of Natural Sciences, Tokyo 181-8588, Japan}
\author[0000-0002-7660-2273]{Xu~Kong}
\affiliation{CAS Key Laboratory for Research in Galaxies and Cosmology, Department of Astronomy, University of Science and Technology of China, Hefei, 230026, China; sfhuang999@ustc.edu.cn, jnac@ustc.edu.cn,twang@ustc.edu.cn}
\affiliation{School of Astronomy and Space Sciences,
University of Science and Technology of China, Hefei, 230026, China}
\affiliation{Institute of Deep Space Sciences, Deep Space Exploration Laboratory, Hefei 230026, China}
\author[0009-0003-5573-9240]{Jiaxun Li}
\affiliation{CAS Key Laboratory for Research in Galaxies and Cosmology, Department of Astronomy, University of Science and Technology of China, Hefei, 230026, China; sfhuang999@ustc.edu.cn, jnac@ustc.edu.cn,twang@ustc.edu.cn}
\affiliation{School of Astronomy and Space Sciences,
University of Science and Technology of China, Hefei, 230026, China}
\author[0000-0002-8391-5980]{Long li}
\affiliation{CAS Key Laboratory for Research in Galaxies and Cosmology, Department of Astronomy, University of Science and Technology of China, Hefei, 230026, China; sfhuang999@ustc.edu.cn, jnac@ustc.edu.cn,twang@ustc.edu.cn}
\affiliation{School of Astronomy and Space Sciences,
University of Science and Technology of China, Hefei, 230026, China}
\author[0000-0003-4156-3793]{Jian-Guo Wang}
\affiliation{Yunnan Observatories, Chinese Academy of Sciences, Kunming 650011, PR China}
\affiliation{Key Laboratory for the Structure and Evolution of Celestial Objects, Yunnan Observatories, Kunming 650011, China}
\author{Ze-Lin Xu}
\affiliation{CAS Key Laboratory for Research in Galaxies and Cosmology, Department of Astronomy, University of Science and Technology of China, Hefei, 230026, China; sfhuang999@ustc.edu.cn, jnac@ustc.edu.cn,twang@ustc.edu.cn}
\affiliation{School of Astronomy and Space Sciences,
University of Science and Technology of China, Hefei, 230026, China}
\author[0000-0002-1935-8104]{Yongquan Xue}
\affiliation{CAS Key Laboratory for Research in Galaxies and Cosmology, Department of Astronomy, University of Science and Technology of China, Hefei, 230026, China; sfhuang999@ustc.edu.cn, jnac@ustc.edu.cn,twang@ustc.edu.cn}
\affiliation{School of Astronomy and Space Sciences,
University of Science and Technology of China, Hefei, 230026, China}
\author[0000-0002-7330-4756]{Ye-Fei Yuan}
\affiliation{CAS Key Laboratory for Research in Galaxies and Cosmology, Department of Astronomy, University of Science and Technology of China, Hefei, 230026, China; sfhuang999@ustc.edu.cn, jnac@ustc.edu.cn,twang@ustc.edu.cn}
\affiliation{School of Astronomy and Space Sciences,
University of Science and Technology of China, Hefei, 230026, China}

\author{Jingquan Cheng}
\affiliation{Purple Mountain Observatory, Chinese Academy of Sciences, Nanjing 210023, China}
\author[0000-0003-4200-4432]{Lulu Fan}
\affiliation{CAS Key Laboratory for Research in Galaxies and Cosmology, Department of Astronomy, University of Science and Technology of China, Hefei, 230026, China; sfhuang999@ustc.edu.cn, jnac@ustc.edu.cn,twang@ustc.edu.cn}
\affiliation{School of Astronomy and Space Sciences,
University of Science and Technology of China, Hefei, 230026, China}
\affiliation{Institute of Deep Space Sciences, Deep Space Exploration Laboratory, Hefei 230026, China}

\author{Jie Gao}
\affiliation{State Key Laboratory of Particle Detection and Electronics, University of Science and Technology of China, Hefei 230026, China}

\author[0000-0001-7201-1938]{Lei Hu}
\affiliation{Purple Mountain Observatory, Chinese Academy of Sciences, Nanjing 210023, China}
\affiliation{McWilliams Center for Cosmology, Department of Physics, Carnegie Mellon University, 5000 Forbes Ave, Pittsburgh, 15213, PA, USA}

\author[0000-0003-3424-3230]{Weida Hu}
\affiliation{Department of Physics, University of California, Santa Barbara, Santa Barbara, CA 93106, USA}

\author{Bin Li}
\affiliation{Purple Mountain Observatory, Chinese Academy of Sciences, Nanjing 210023, China}

\author{Feng Li}
\affiliation{State Key Laboratory of Particle Detection and Electronics, University of Science and Technology of China, Hefei 230026, China}
\author{Ming Liang}
\affiliation{National Optical Astronomy Observatory (NSF's National Optical-Infrared Astronomy Research Laboratory) 950 N Cherry Ave. Tucson Arizona 85726, USA}

\author{Hao Liu}
\affiliation{State Key Laboratory of Particle Detection and Electronics, University of Science and Technology of China, Hefei 230026, China}

\author{Wei Liu}
\affiliation{Purple Mountain Observatory, Chinese Academy of Sciences, Nanjing 210023, China}
\author{Zheng Lou}
\affiliation{Purple Mountain Observatory, Chinese Academy of Sciences, Nanjing 210023, China}
\author[0000-0003-1297-6142]{Wentao Luo}
\affiliation{Institute of Deep Space Sciences, Deep Space Exploration Laboratory, Hefei 230026, China}
\author{Yuan Qian}
\affiliation{Purple Mountain Observatory, Chinese Academy of Sciences, Nanjing 210023, China}

\author{Jinlong Tang}
\affiliation{Institute of Optics and Electronics, Chinese Academy of Sciences, Chengdu 610209, China}
\author[0000-0002-3105-3821]{Zhen Wan}
\affiliation{CAS Key Laboratory for Research in Galaxies and Cosmology, Department of Astronomy, University of Science and Technology of China, Hefei, 230026, China; sfhuang999@ustc.edu.cn, jnac@ustc.edu.cn,twang@ustc.edu.cn}
\affiliation{School of Astronomy and Space Sciences,
University of Science and Technology of China, Hefei, 230026, China}
\author{Hairen Wang}
\affiliation{Purple Mountain Observatory, Chinese Academy of Sciences, Nanjing 210023, China}
\author[0000-0003-1617-2002]{Jian Wang}
\affiliation{State Key Laboratory of Particle Detection and Electronics, University of Science and Technology of China, Hefei 230026, China}
\affiliation{Institute of Deep Space Sciences, Deep Space Exploration Laboratory, Hefei 230026, China}

\author[0000-0001-7768-7320]{Ji Yang}
\affiliation{Purple Mountain Observatory, Chinese Academy of Sciences, Nanjing 210023, China}
\author{Dazhi Yao}
\affiliation{Purple Mountain Observatory, Chinese Academy of Sciences, Nanjing 210023, China}

\author[0000-0002-1463-9070]{Hongfei Zhang}
\affiliation{State Key Laboratory of Particle Detection and Electronics, University of Science and Technology of China, Hefei 230026, China}

\author{Xiaoling Zhang}
\affiliation{Purple Mountain Observatory, Chinese Academy of Sciences, Nanjing 210023, China}

\author[0000-0002-1330-2329]{Wen Zhao}
\affiliation{CAS Key Laboratory for Research in Galaxies and Cosmology, Department of Astronomy, University of Science and Technology of China, Hefei, 230026, China; sfhuang999@ustc.edu.cn, jnac@ustc.edu.cn,twang@ustc.edu.cn}
\affiliation{School of Astronomy and Space Sciences,
University of Science and Technology of China, Hefei, 230026, China}

\author[0000-0003-3728-9912]{Xianzhong Zheng}
\affiliation{Purple Mountain Observatory, Chinese Academy of Sciences, Nanjing 210023, China}

\author[0000-0003-0694-8946]{Qingfeng Zhu}
\affiliation{CAS Key Laboratory for Research in Galaxies and Cosmology, Department of Astronomy, University of Science and Technology of China, Hefei, 230026, China; sfhuang999@ustc.edu.cn, jnac@ustc.edu.cn,twang@ustc.edu.cn}
\affiliation{School of Astronomy and Space Sciences,
University of Science and Technology of China, Hefei, 230026, China}
\affiliation{Institute of Deep Space Sciences, Deep Space Exploration Laboratory, Hefei 230026, China}

\author{Yingxi Zuo}
\affiliation{Purple Mountain Observatory, Chinese Academy of Sciences, Nanjing 210023, China}

\begin{abstract}
High-cadence, multiwavelength observations have continuously revealed the diversity of tidal disruption events (TDEs), thus greatly advancing our knowledge and understanding of TDEs. In this work, we conducted an intensive optical-UV and X-ray follow-up campaign of TDE AT~2023lli and found a remarkable month-long bump in its UV/optical light curve nearly two months prior to maximum brightness. The bump represents the longest separation time from the main peak among known TDEs to date. The main UV/optical outburst declines as $t^{-4.10}$, making it one of the fastest- decaying optically selected TDEs. Furthermore, we detected sporadic X-ray emission 30 days after the UV/optical peak, accompanied by a reduction in the period of inactivity. It is proposed that the UV/optical bump could be caused by the self-intersection of the stream debris, whereas the primary peak is generated by the reprocessed emission of the accretion process. In addition, our results suggest that episodic X-ray radiation during the initial phase of decline may be due to the patched obscurer surrounding the accretion disk, a phenomenon associated with the inhomogeneous reprocessing process. The double TDE scenario, in which two stars are disrupted in sequence, is also a possible explanation for producing the observed early bump and main peak. We anticipate that the multicolor light curves of TDEs, especially in the very early stages, and the underlying physics can be better understood in the near future with the assistance of dedicated surveys such as the deep high-cadence survey of the 2.5 m Wide Field Survey Telescope.

\end{abstract}


\section{Introduction} \label{sec:intro}
A tidal disruption event (TDE) occurs when a star gets too close to the supermassive black hole (SMBH) at the center of a galaxy and is torn apart by its gravitational force. In recent years, many TDEs have been detected by sky survey projects using various telescopes and instruments \citep{Velzen2019,Velzen2021,Gezari2021,Jiang2021,Wang2022,Hammerstein2023a,Yao2023,Zhu2023}. These TDEs commonly show a blue continuum that can be characterized as a blackbody of nearly constant temperature ($\sim 10^4~\rm K$) over time, broad emission lines (such as H$\alpha$ and He \textsc{ii}) and a ``fast rise and slow decay'' pattern \citep{Arcavi2014,Velzen2020,Gezari2021}.

Most TDEs that have been discovered so far exhibit a smooth single-peak light curve in both UV and optical bands. However, some of them display a bump in the rising phase of their light curves \citep[][Lin et al. in preparation.]{Holoien2019,Charalampopoulos2023,Faris2023,Huang2023b,Wangyb2023}. 
This early signature may be produced by a number of physical processes, such as the cooling of unbounded debris, vertical compression during the first pericenter passage, shock breakout of self-crossing debris and so-called ``TDE encore''~\citep{Kasen2010,Yalinewich2019,Wangyb2023,Ryu2024}. But no definitive conclusion has been reached.
The collision between stream debris and the disk was proposed to explain the early bump in recent UV/optical outbursts of the candidate for repeated partial TDE (pTDE) ASASSN-14ko~\citep{Huang2023b}. Furthermore, a fraction of the sources appeared to rebrighten in the optical light curve as the duration of observation increased~\citep{Jiang2019,Malyali2021,Chen2022,Wang2023,Yao2023}. Rebrightening in TDEs has also been found in X-rays, and the likely source is a repeated pTDE~\citep{Liu2023,Liu2024,Miniutti2023,Wevers2023}. 
A repeated pTDE has been applied to explain the periodic optical outburst of ASASSN-14ko \citep{Payne2021} and the optical rebrightening in AT 2020vdq \citep{Somalwar2023}.

The source of the X-ray emission is unquestionably the accretion processes, but there is an ongoing debate regarding the origin of the UV/optical emissions. Two rival models are being discussed: one suggests that UV/optical emission is a result of the reprocessing of X-ray photons in an extended envelope/outflow \citep{Loeb1997,Strubbe2009,Metzger2016,Roth2016,Dai2018}, while the other proposes that the emissions are caused by shocks generated by the stream-stream collision of the stellar debris \citep{Piran2015,Jiang2016,Huangxiaoshan2023,Jankovivc2023}.
An intriguing phenomenon in optical TDEs is that X-ray emission lags UV/optical emission \citep{Gezari2017,Pasham2017,Shu2020,Velzen2021,Chen2022,Liu2022,Wang2022,Wang2023,Huang2023a}. This can be explained by a structural change in the reprocessing layer as the accretion rate declines from super-Eddington to sub-Eddington (e.g. \citealt{Thomsen2022}) or by a delayed onset of accretion after circularization. 

AT~2023lli (RA=22:57:39.470, DEC=+40:32:40.02) is a TDE discovered by the Gravitational-wave Optical Transient Observer (GOTO) on 2023 Jun 23 at a redshift of 0.036\footnote{https://www.wis-tns.org/object/2023lli}. We conducted multiwavelength follow-up observations of this source and found that it showed a strong UV/optical bump and a delayed and episodic X-ray emission. In Section \ref{sec:data}, we describe the observations and data analysis methods, and the multiwavelength results are represented in Section \ref{sec:results}. In Section \ref{sec:discussion}, we discuss the possible physical mechanisms for the early bump and the X-ray emission. In Section \ref{sec:conclusion}, we summarize our main results and conclusions. In this work, we assume cosmological parameters of $H_0=70\,\text{km}\,\text{s}^{-1}\,\text{Mpc}^{-1}$, $\Omega_{\rm M}=0.3$, and $\Omega_{\Lambda}=0.7$.

\begin{figure*}
    \centering
    \includegraphics[width=0.75\textwidth]{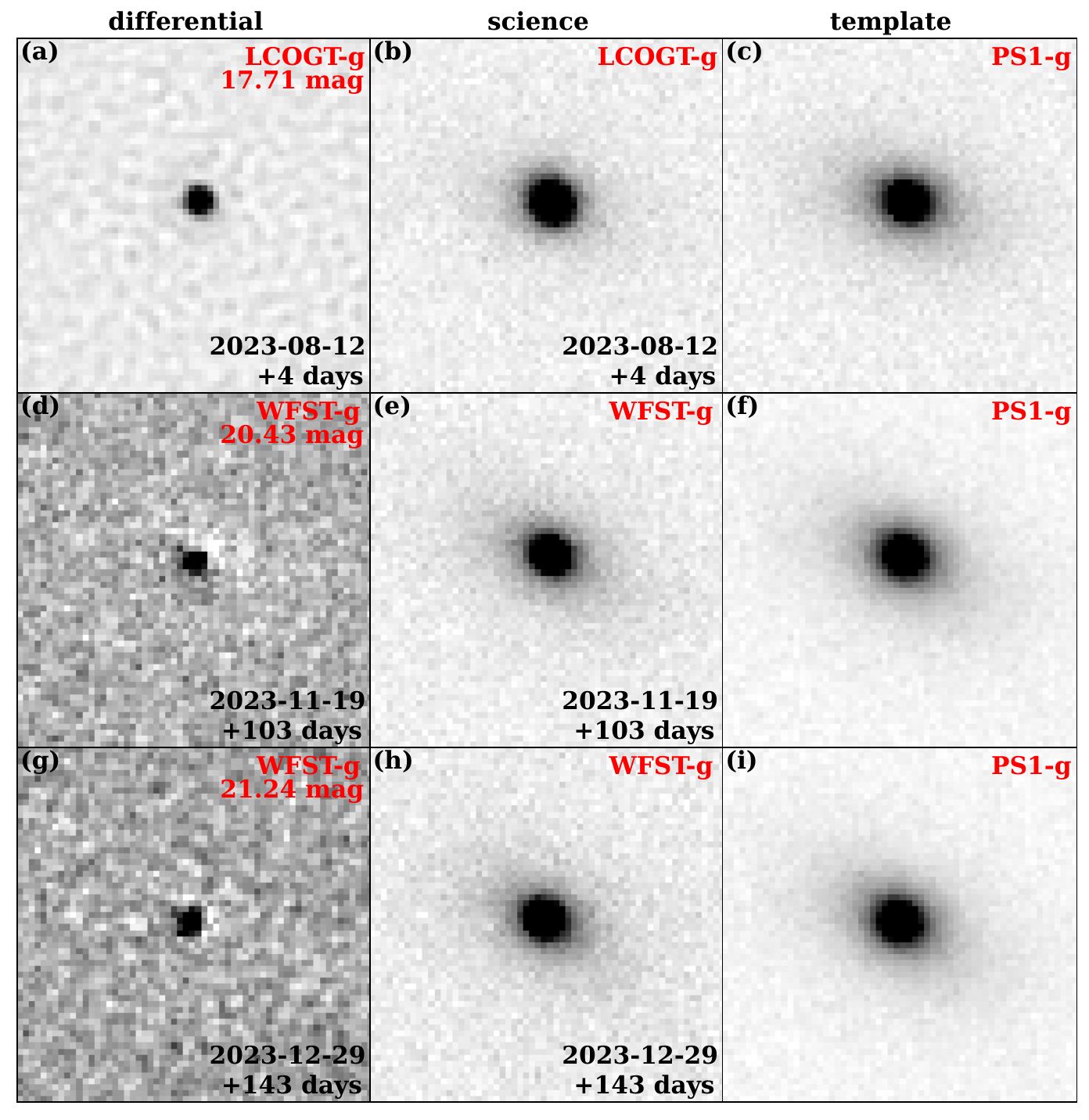}
    \caption{The differential, science and template images for LCOGT and WFST are shown. Panels (a) and (b) show the observations on MJD 60168, when the source was close to its UV/optical peak. Panels (d) and (e) are the images observed on MJD 60267, and panels (g) and (h) are the images observed on MJD 60307. The UV/optical emission was faint in both observations. The indicated magnitudes are measured directly from the differential images without correction for Galactic extinction. The observed time and the day relative to the optical peak are marked in the bottom right on the first and second columns. }
    \label{fig:image}
\end{figure*}

\section{Observations and Data Reduction} \label{sec:data}
\subsection{Swift X-Ray Photometry}
We proposed several target-of-opportunity (ToO) observations to monitor the multiwavelength evolution using the X-Ray Telescope (XRT; \citealt{Burrows2005}) and the Ultraviolet/Optical Telescope (UVOT; \citealt{Roming2005}) on board the Swift Observatory. We also retrieved the public data (PI: Hinkle) from the High Energy Astrophysics Science Archive Research Center website. Swift data were processed with HEASoft 6.32.1. We ran the tasks \texttt{xrtpipeline} and \texttt{xrtproducts} to generate light curves and spectra. The source and background count rates were extracted from circular regions centered on the object with a radius of $47^{\prime\prime}.1$ and a nearby circular region with a radius of $150^{\prime\prime}$, respectively. Source detection was performed using the \texttt{ximage} task. For data in which the source was not detected, we derived a 3$\sigma$ upper limit on the X-ray flux using the WebPIMMS tool\footnote{https://heasarc.gsfc.nasa.gov/cgi-bin/Tools/w3pimms/w3pimms.pl} and assuming a power-law X-ray spectrum with a photon index of 1.75 \citep{Ricci2017}. To improve the signal-to-noise ratio (S/N), we stacked the event files for MJD 60198-60201 (ObsIDs: 00016100029 and 00016100030) and MJD 60226-60244 (ObsIDs: 00016100038, 00016100039, and 00016100040), respectively. In these two stacked files, the source was detected above the 3$\sigma$ level using the \texttt{ximage} task.

\subsection{XMM-Newton Observation}
We proposed a ToO observation with XMM-Newton, which was performed on MJD 60295.5 (ObsID: 0932391301) with a total observation time of 20 ks. The XMM-Newton data were reduced with the Science Analysis System (version 21.0.0). The tasks \texttt{cifbuild} and \texttt{odfingest} were executed for data preparation, and then light curves and spectra were extracted through the task \texttt{xmmextractor}.

\subsection{Swift UV/Optical Photometry}
We ran \texttt{uvotimsum} to sum up the images and then generate the light curves through the task \texttt{uvotsource} with the source and background regions defined by circles with radii of $10^{\prime\prime}$ and $30^{\prime\prime}$, respectively.  

\begin{figure*}
    \centering
\includegraphics[width=0.9\textwidth]{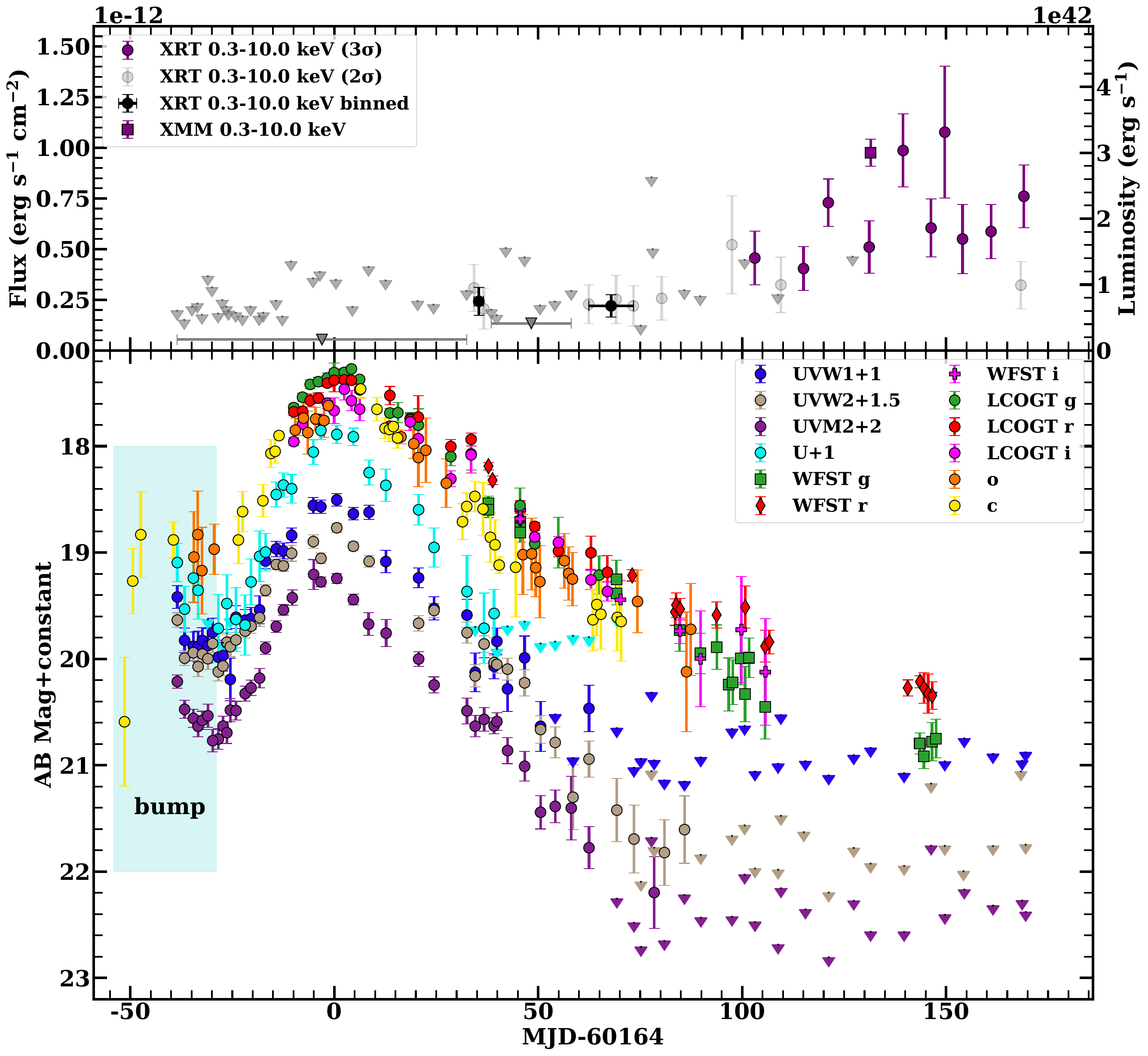}
    \caption{Multiwavlength light curves of AT~2023lli. The top panel shows the Swift/XRT X-ray fluxes (or upper limits) in the 0.3--10.0 keV band, with 2$\sigma$ detections shown as grey dots and 3$\sigma$ detections as purple dots. Additionally, a single epoch of XMM-Newton observation is overplotted in the purple square. The binned data are presented in black, with horizontal error bars indicating the time range. The bottom panel displays the UV/optical light curves obtained from Swift/UVOT, ATLAS, LCOGT, and WFST, with a shadowed region indicating a bump in the rising stage. In the bottom panel, different bands and the magnitude offset are labeled. All data have been corrected for Galactic extinction. Triangles represent the 3$\sigma$ upper limits.}
    \label{fig:lc_mw}
\end{figure*}

\subsection{Ground-based Optical Photometry}

AT~2023lli was first detected by the Asteroid Terrestrial impact Last Alert System (ATLAS; \citealt{ATLAS,Smith2020}). The ATLAS $c$- and $o$-band light curves were obtained using the ATLAS Forced Photometry Service, which produces point spread function (PSF) photometry on the difference images. ATLAS has three to four single exposures within each epoch (typically within one day), so we binned the light curve every epoch to improve the S/N.

Upon noticing that AT~2023lli displayed a rebrightening after the initial peak (the ``early bump''), we promptly conducted optical monitoring in the $gri$ bands using 1.0 m telescopes from the Las Cumbres Observatory Global Telescope (LCOGT) network \citep{LCOGT} starting from 2023 July 28. The observation continued until October 18, when it almost faded out in the LCOGT exposures. The observations were carried out every 2--3 days with a denser sampling around the peak, using exposure times of 260s, 160s, and 260s in the $g$, $r$, and $i$ bands, respectively.

Subsequently, we also initiated a follow-up campaign with the 2.5 m
Wide Field Survey Telescope (WFST) in the $gri$ bands with an exposure time of 30 s from 2023 September 14. There is a gap between November 21 and December 29 due to engineering maintenance and testing. The exposure time was changed to 60 s after the observations were resumed. Briefly, the WFST is a newly established photometric facility equipped with a 2.5 m diameter primary mirror, an active optics system, and a mosaic CCD camera with 0.73 gigapixels on the primary focal plane, allowing high-quality image capture over a 6.5 $\rm deg^2$ field of view~\citep{WFST}. It was installed near the summit of Saishiteng Mountain in northwestern China in the summer of 2023 and entered its commissioning stage in September following the official release of the first-light image of the Andromeda galaxy\footnote{https://wfst.ustc.edu.cn/news-and-meetings/news/20230917/20230917/}. Follow-up observations with the WFST are particularly important when AT~2023lli becomes fainter, making its detection challenging with the 1m LCOGT network.

We used PanSTARRS (\citealt{PS1}) $gri$ band stack images as reference images and employed {\tt HOTPANTS}\,\citep{Becker2015} for image subtraction. Prior to subtraction, we removed cosmic rays and aligned the images using Astrometry.net. After image subtraction, we performed PSF photometry on the difference image using the Photutils package of Astropy \citep{Astropy} for the $gri$ data. We show representative images at different stages (magnitudes) in Figure~\ref{fig:image}. Galactic extinction is corrected for $E(B-V)=0.118$, derived using an online tool\footnote{\url{https://irsa.ipac.caltech.edu/applications/DUST/}}, and the extinction law of \cite{Cardelli1989}. The multiwavelength light curves are presented in Figure~\ref{fig:lc_mw}.

\begin{figure*}
    \centering
    \includegraphics[width=1.0\textwidth]{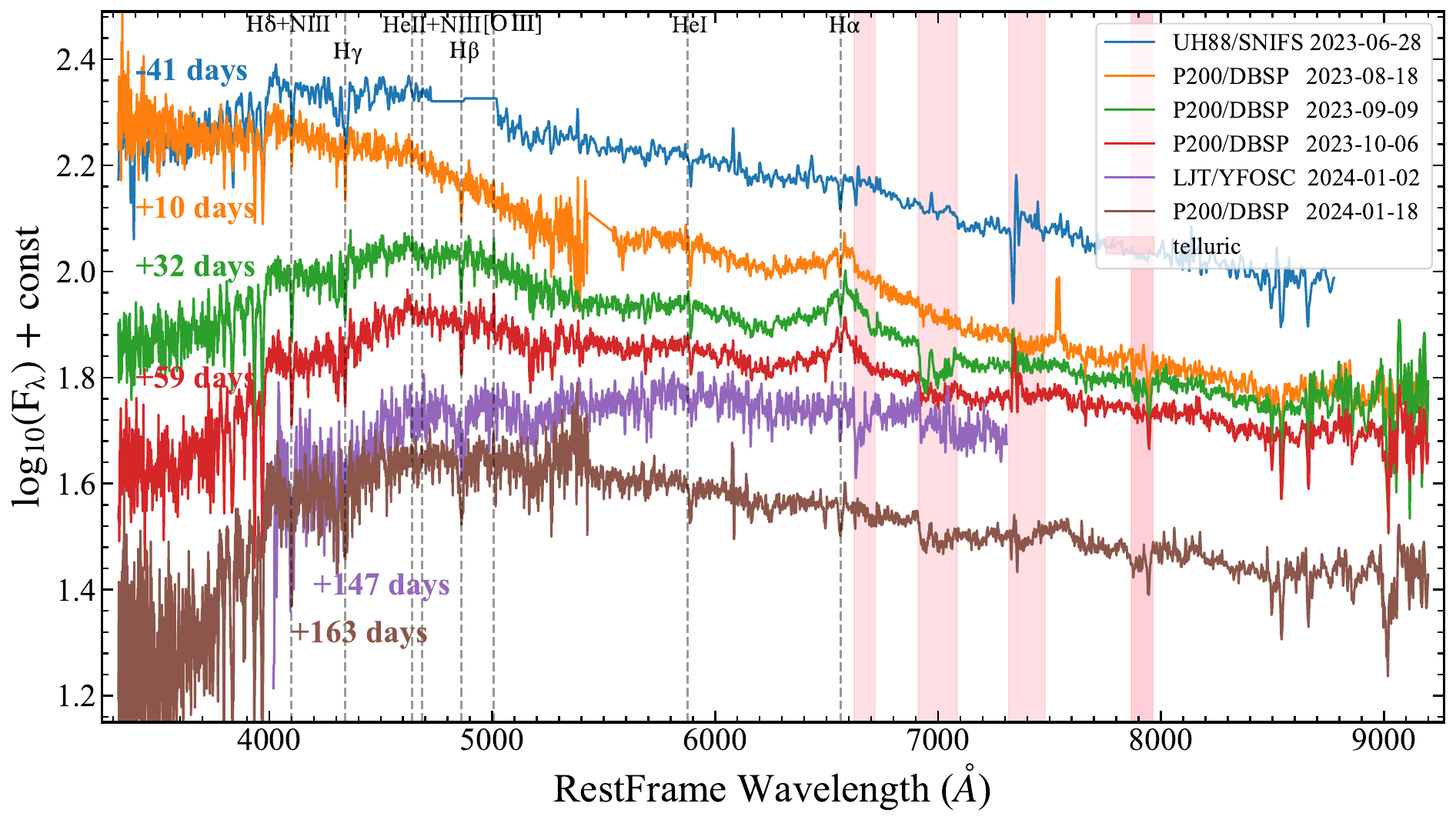}
    \caption{The Galactic-extinction-corrected spectra for AT~2023lli. The pink regions highlight the prominent telluric regions. The gray dashed lines mark some characteristic emission lines. The phases relative to the peak were marked on the left side with the same color of the corresponding spectrum. }
    \label{fig:all_spec}
\end{figure*}

\subsection{Spectral Observations and Data Reduction} 

Shortly after its discovery, the SCAT group \citep{Tucker2022} acquired a spectrum of AT~2023lli utilizing the Supernova Integral Field Spectrograph, which was installed on the UH88 telescope \citep{Hinkle2023}. This spectrum was retrieved from the Transient Name Server (TNS) website\footnote{https://www.wis-tns.org/object/2023lli}. We then acquired four more spectra using the Double Spectrograph (DBSP) instrument mounted on the Hale 200 inch telescope at Palomar Observatory (P200, \citealt{Oke1982}). For these observations, the D55 dichroic was adopted to split the incoming light longer than or shorter than 5500~\AA~into separate red and blue channels. A grism of 600 lines $\rm mm^{-1}$ blazed at 3780~\AA~was used for the blue arm, and a grism of 316 lines $\rm mm^{-1}$ blazed at 7150~\AA~was used for the red arm. The data were then reduced using the Python package \texttt{Pypeit} (\citealt{Pypeit2,Pypeit1}), which can fully automatically implement the standard reduction procedure for long-slit spectroscopic observations. Furthermore, we obtained another spectrum using the Yunnan Faint Object Spectrograph and Camera (YFOSC) on board the LiJiang 2.4 m telescope~\citep{Fan2015-2m4,Wang2019-2m4} on 2024 January 2. All spectra are shown in Figure~\ref{fig:all_spec}. Broad $\rm H\alpha$ ($\sim$10,000 km~s$^{-1}$) and probable Helium lines are detected when the source is bright. As the TDE emission significantly diminished, the spectrum (+147 and +1365 days) is dominated by host galaxy emission, showing deep Balmer absorption (see Figure~\ref{fig:all_spec}), consistent with the types of galaxies that TDEs are known to preferentially inhabit~\citep{French2016}.

\section{Results}\label{sec:results}
\subsection{Multiwavelength Light Curves}

The multiwavelength light curves of AT~2023lli show a complex structure with an early bump and a main peak in UV/optical bands. The source was first detected by GOTO on MJD 60118, with a magnitude of 18.84 mag in the L band \citep{Hinkle2023}. We followed up the initial observation by ATLAS and found that the source had already been visible on MJD 60112.57 with a magnitude of $20.59\pm{0.61}$ mag in the c band. The source was then brightened to $18.83\pm{0.4}$ mag in the same band by MJD 60116.56, before fading. We missed the early UV emission of the source. The first UVOT observation was made on MJD 60125.5, when the source was already declining, with host-subtracted magnitudes of $18.42\pm{0.11}$, $18.21\pm{0.06}$ and $18.14\pm{0.07}$ mag for the \textsl{UVW1}, \textsl{UVM2} and \textsl{UVW2} bands, respectively. The source then rose again after MJD 60140 and reached the optical peak on MJD 60164, with the magnitudes of $17.31\pm{0.09}$, $17.38\pm{0.11}$ and $17.67\pm{0.12}$ mag for the \textsl{g}, \textsl{r} and \textsl{i} bands, respectively. At that time, we obtained magnitudes of $17.51\pm{0.06}$, $17.24\pm{0.05}$, $17.27\pm{0.04}$, and $17.39\pm{0.09}$ mag for the \textsl{UVW1}, \textsl{UVM2}, \textsl{UVW2} and \textsl{U} bands, respectively.

We stacked a total of 27 Swift/XRT event files with an effective exposure time of 59.1 ks, covering the period from MJD 60125 to 60188. From these stacked data, we obtained a 3$\sigma$ upper limit of 0.0011 $\rm counts\,s^{-1}$ corresponding to a luminosity of $\rm 1.66\times 10^{41}\,erg\,s^{-1}$ by fixing the Galactic hydrogen column density value of $9.05\times 10^{20}\,\text{cm}^{-2}$ \citep{HI4PI2016} and assuming a photon index of 1.75 \citep{Ricci2017}. The source was initially detected in the X-ray band until MJD 60198, where a 2.5$\sigma$ detection was achieved with a count rate of $0.0058\pm{0.0022}$ $\rm counts\,s^{-1}$. Additionally, the source was also detected on MJD 60200. To improve the S/N, we combined the event files from two ObsIDs (00016100029 and 00016100030), which covered the time period from MJD 60198 to 60200. From these stacked data, we derived a mean count rate of $0.0045\pm{0.001}$ $\rm counts\,s^{-1}$ at the 3$\sigma$ level, with an effective exposure time of 4013 s. However, in a subsequent observation with Swift/XRT lasting 4152 s, only a 3$\sigma$ upper limit of 0.004 $\rm counts\,s^{-1}$ was obtained. By merging three event files from MJD 60226 to 60237 (ObsIDs: 00016100038--00016100040), we derived a mean count rate of $0.004\pm{0.001}$ $\rm counts\,s^{-1}$. In the following Swift/XRT observation with an exposure time of 3678 s, the source was not detected, and a 3$\sigma$ upper limit of 0.003 $\rm counts\,s^{-1}$ is imposed. Thus, the X-ray emission from the source appeared to exhibit a sporadic behavior with a general upward trend. At MJD 60295.5, XMM-Newton recorded an unabsorbed luminosity of $\rm (3.01\pm{0.21})\times 10^{42}~erg~s^{-1}$ in the X-ray. Following this, the X-ray emission was detectable in every subsequent observation with XRT. When the X-ray showed sporadic behavior, we checked the data and found that the count rate was above the statistical fluctuation and confirmed that such behavior actually existed (P-value of 0.0004). For XRT, the conversion factor from count rate to flux was determined to be $5.30\times 10^{-11} \rm erg~cm^{-2} counts^{-1}$, which was derived from the spectral fit of the stacked spectrum extracted from all event files with X-ray detection.

\begin{figure}
    \centering
    \includegraphics[width=0.47\textwidth]{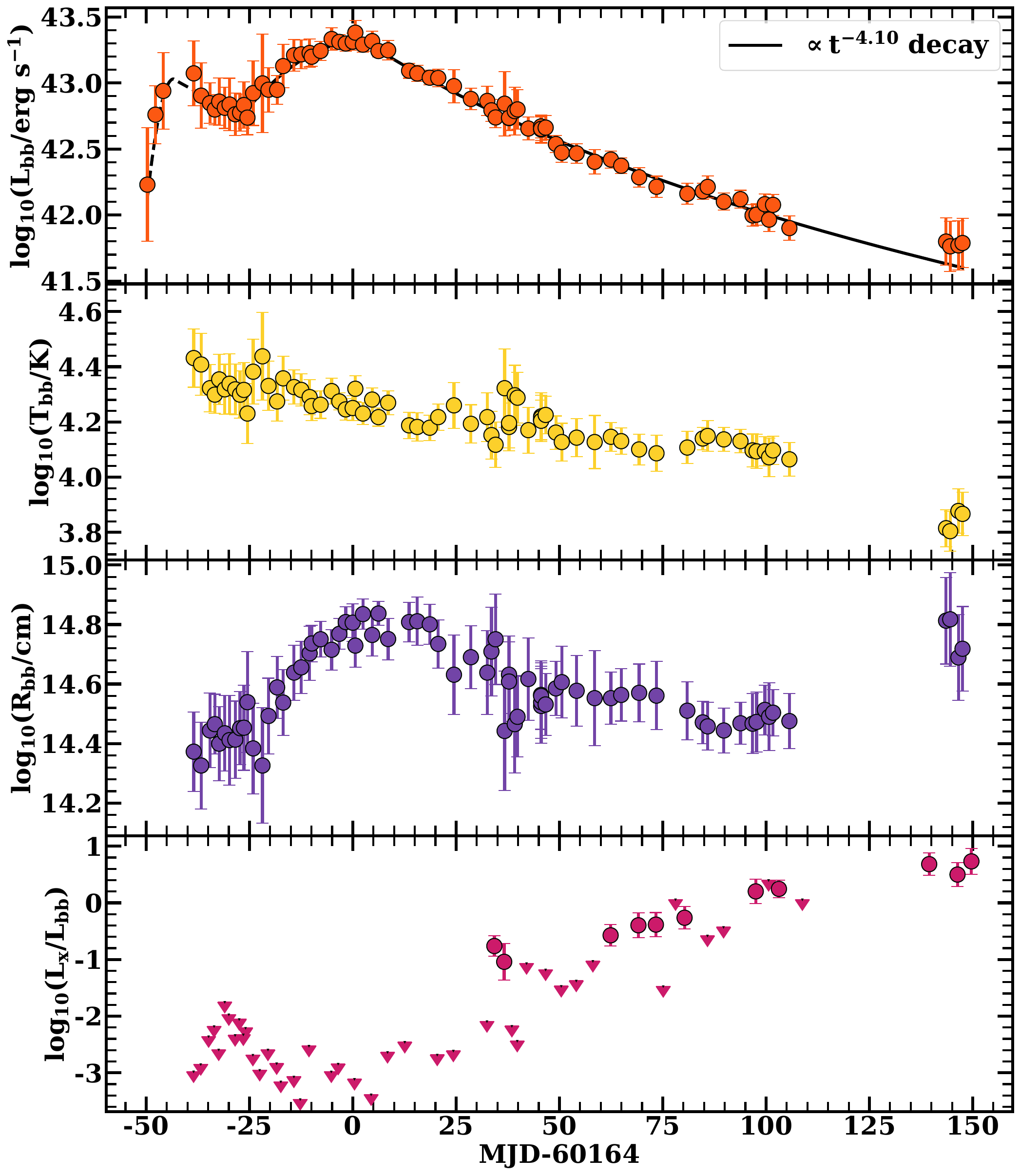}
    \caption{The top panel displays the UV/optical black body luminosity of AT~2023lli, with the black line representing the fitting by a ``Gaussian rise and power-law decay'' model. The fitted results for the bump and main outburst are represented by dashed and solid lines, respectively. The main peak declined following $t^{-4.10}$. The second and third panels illustrate the blackbody temperature and radius, respectively. The bottom panel shows the ratio of X-ray luminosity to UV-optical blackbody luminosity. Triangles represent the 3$\sigma$ upper limits.}
    \label{fig:lc_bb}
\end{figure}

The UV/optical SEDs were fitted with a blackbody model and \texttt{SuperBol} \citep{Nicholl2018} was used to calculate the temporal evolution of the blackbody luminosity ($L_{\rm bb}$), temperature ($T_{\rm bb}$) and blackbody radius ($R_{\rm bb}$). The results are displayed in Figure~\ref{fig:lc_bb}. The highest $L_{\rm bb}$ was recorded on MJD 60164.6 with a value of $(2.40\pm{0.51})\times10^{43}\,\rm erg\,s^{-1}$ and at that time we derived $T_{\rm bb}=(2.09\pm{0.23})\times 10^{4}\,\rm K$ and $R_{\rm bb}=(5.36\pm{0.90})\times 10^{14}\,\rm cm$. After that, the luminosity decreases, while no significant temporal evolution can be seen in the blackbody temperature. We derived the energy released in the UV/optical bands of $\rm 1.6\times 10^{49} erg$ and $\rm 8.8\times 10^{49} erg$ during the bump and the main outburst, respectively. Using the \texttt{emcee} python package \citep{Foreman2013}, we employed a ``Gaussian rise and power-law decay'' model to fit $L_{\rm bb}$ in both the bump and the main outburst, as described in \cite{Velzen2019,Velzen2021,Hammerstein2023a,Yao2023}. We obtained a power-law index of -4.10, which is much steeper than the theoretical prediction of -5/3 \citep{Rees1988}. Moreover, some researchers found that TDEs exhibit a wide range of power-law indices, from -4.0 to -0.5 \citep{Velzen2019,Velzen2021,Gezari2021,Hammerstein2023a}. Our result resembles that of AT~2019mha, the most rapidly fading source (with a power-law index of 4.0) among the optically selected TDEs studied by \cite{Velzen2021}. This makes AT~2023lli one of the fastest declining TDEs ever observed.  A steep slope in the light curve has been predicted by some theoretical work \citep{Guillochon2013,Ryu2020} that involves pTDEs. As showen by \cite{Ryu2020}, a more rapid decrease in light curves is associated with a lower amount of mass being removed from the star. Consequently, the situation of AT~2023lli could be connected to a pTDE. A recent systematic analysis conducted on optically-selected TDEs has found that the ratio between UV/optical and X-ray luminosities ($L_{\rm bb}/L_{\rm X}$) tends to approach values similar to those of a disk ($0.5-10$) at later times, although it exhibits a wide range of values at earlier times~\citep{Guolo2023}. This trend is also observed in the case of AT~2023lli. As the X-ray emission becomes visible and increases in brightness, the ratio $L_{\rm X}/L_{\rm bb}$ gradually increases, with the most recent value slightly above 1 (refer to the bottom panel of Figure~\ref{fig:lc_bb}).

\subsection{X-Ray Spectral Fitting Results}
To improve the S/N ratio of the Swift/XRT spectrum, we combined nine event files (ObsID: 00016100029, 00016100030, 00016100038--00016100040, 00016100043, 00016100047, 00016100049, 00016100051) with a total exposure time of 17.27 ks. The spectrum is extracted using the \texttt{xselect} tool and fitted with the \texttt{tbabs*zashift*powerlaw} model. The absorption column density is fixed at the Galactic value of $9.05\times 10^{20}\,\text{cm}^{-2}$ \citep{HI4PI2016}. This fit results in a photon index of $\Gamma=4.15_{-0.58}^{+0.64}$ (Cstat/dof=22.73/22) using \texttt{xspec v12.13.1}. A soft X-ray spectrum in TDEs is usually believed to originate from disk emission. Alternatively, the spectrum can be equally well fitted with the \texttt{tbabs*zashift*diskbb} model, yielding a temperature of $k \rm T_{\rm in}=0.146_{-0.024}^{+0.031}~keV$ (Cstat/dof=23.26/22).

\begin{figure}
    \centering
    \includegraphics[width=0.45\textwidth]{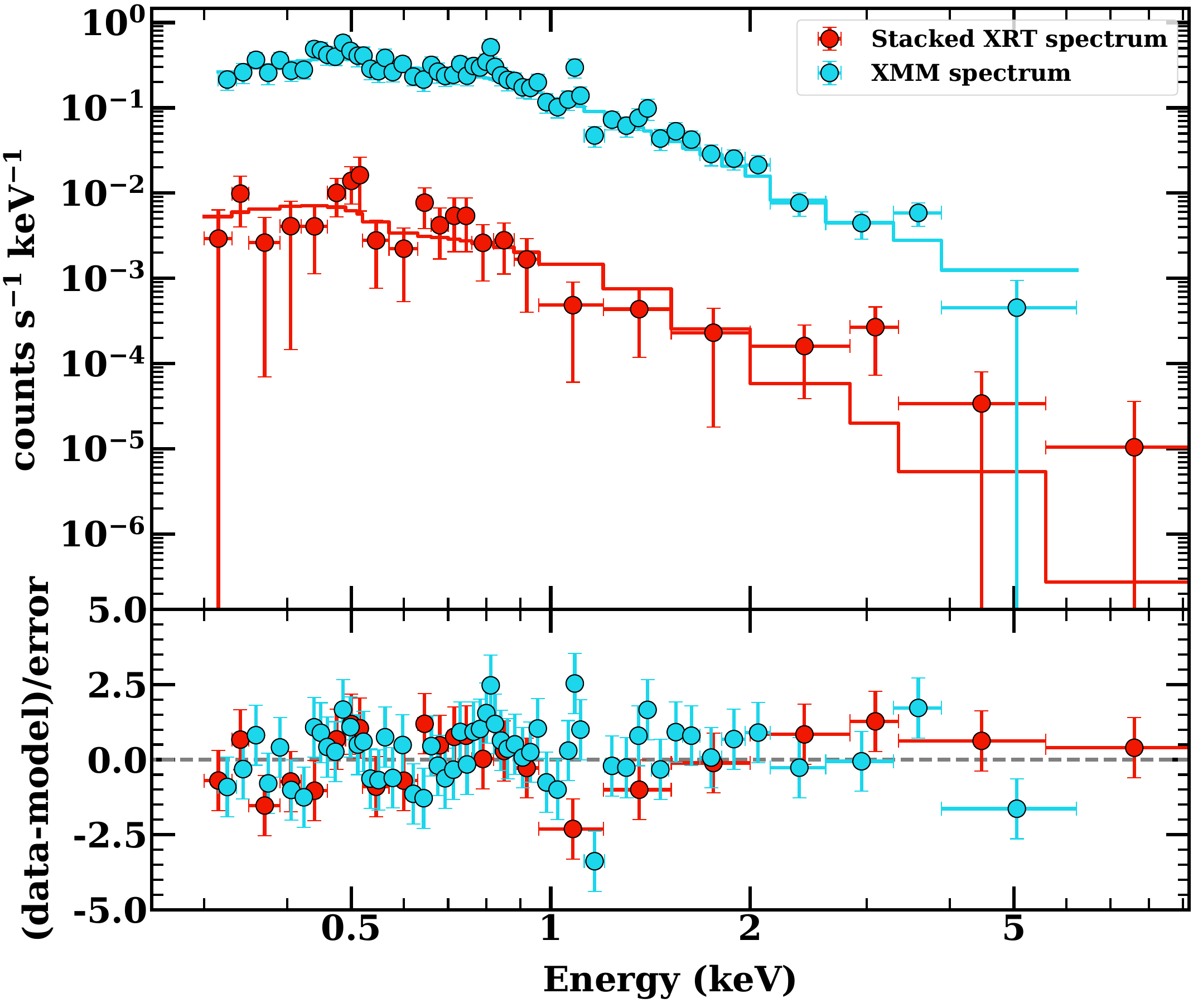}
    \caption{The fitted results of the XMM-Newton spectrum by an absorbed multicolor disk plus a power-law model (\texttt{tbabs*zashift*(diskbb+powerlaw)}). The XRT spectrum was fitted by an absorbed power-law model (\texttt{tbabs*zashift*powerlaw}).}
    \label{fig:xspec}
\end{figure}

To utilize $\chi^2$ statistics, the XMM-Newton EPIC-pn spectrum is binned with a minimum of 15 photons. However, a single multicolor disk or blackbody model does not provide a satisfactory fit, leading to a large $\chi^2$ value. Next, we fit the spectrum using an absorbed power-law model,  \texttt{tbabs*zashift*powerlaw}. The fit converges to an index $3.16^{+0.12}_{-0.11}$ ($\chi^2$/dof=72.15/57), indicating a soft X-ray spectrum. However, this model did not account for the possible thermal emission from the disk. Therefore, we tried another model \texttt{tbabs*zashift*(diskbb+powerlaw)}. This model produced a disk temperature of $0.167^{+0.025}_{-0.026}~\rm keV$ and a photon index of $2.63^{+0.38}_{-0.41}$ with $\chi^2$/dof=64.61/55.  Taking into account the potential intrinsic absorption, we tried to add a photoelectric absorbed component with \texttt{tbabs*zashift*wabs*(diskbb+powerlaw)} in the fitting. The parameters derived for this model were a disk temperature of $0.111^{+0.047}_{-0.028}~\rm keV$, a photon index of $2.96^{+0.54}_{-0.48}$, and a hydrogen density of $1.41_{-1.23}^{+2.06}\times 10^{21}\,\text{cm}^{-2}$ ($\chi^2$/dof=60.57/54). The power-law component could originate from nonthermal emission from the corona above the disk. To determine whether the absorbed component is necessary, we performed the F-test on the results and derived the F value of 3.6 and the P-value of 0.06. Therefore, we need to be careful about whether or not we need to include this absorbed component. The result is depicted in Figure~\ref{fig:xspec}.
Although both a power-law model and \texttt{diskbb} model can fit the XRT spectrum well, the latter appears to be more plausible for TDEs because the soft X-ray emission in TDEs likely comes from the disk, which can be better characterized by \texttt{diskbb}. For the XMM spectrum, the additive power-law component may suggest the presence of a corona.

\subsection{Host Galaxy Properties}

\begin{figure}
    \centering
    \includegraphics[width=0.47\textwidth]{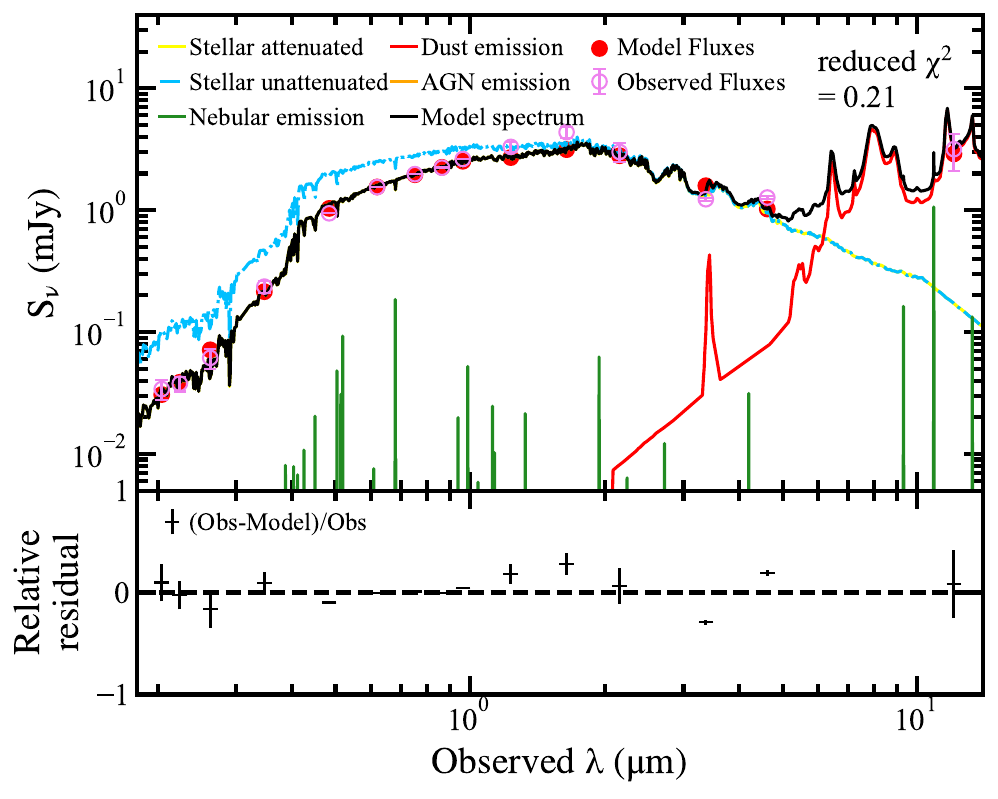}
    \caption{The host SED fitting results using the package CIGALE.
    Top panel: The different components considered in the SED fitting. The red points represent the model flux for each band derived from the best-fit SED, and the violet circles indicate the observed data. It is noteworthy that the AGN contribution to the best-fit SED is zero. Bottom panel: The residuals between the observed data and model flux.}
    \label{host}
\end{figure}

We collected multiband photometry for the host galaxy of AT~2023lli from several archives, including Swift/UVOT, PanSTARRS, the Two Micron All Sky Survey, and Wide field Infrared Survey Explorer. We used the Python package Code Investigating GALaxy Emission (CIGALE; \citealt{CIGALE}) to model the spectral energy distribution (SED) of the host galaxy. CIGALE can fit the SED of a galaxy from far-UV to radio and estimate its physical properties through the analysis of likelihood distribution. We assumed a delayed star formation history with an optional exponential burst and used the single stellar population of \cite{BC03}. We adopted the dust attenuation curve of \cite{Calzetti2000}. Additionally, dust emission is modeled using \cite{Dale2014}, and active galactic nucleus (AGN) emission is calculated with the model of \cite{Stalevski2012,Stalevski2016}. The fitting result is shown in Figure~\ref{host}, which is well fitted solely by stellar components with zero contribution from the AGN component.

Meanwhile, the host galaxy properties were estimated by the CIGALE SED fitting. The stellar mass of the galaxy is $\rm 10^{10.00\pm0.21}\,M_\odot$ and the star formation rate (SFR) is $\rm logSFR = -1.67\pm0.80$. Using the empirical relation between \mbh\ and the total stellar mass of the galaxy in the local universe (see Equation (4) in \citealt{Reines2015}), the central \mbh\ of AT~2023lli is estimated to be $10^{6.40\pm0.47}M_{\odot}$.  We derived the value of u-r of 2.08, which also falls within the ``green valley'', which dominates the host galaxies of TDEs \citep{Hammerstein2021}.

\section{Discussions}\label{sec:discussion}
AT~2023lli is a nuclear transient with broad H$\alpha$ ($\sim$ 10,000 km/s) and possible He~II emission lines. It displays a sudden rise and a slow decay in the UV/optical bands, with a consistently high blackbody temperature maintained over a span of 100 days. Additionally, it exhibits soft X-rays with thermal emission, with a delay relative to the UV/optical bands. These features are typical of TDEs. However, AT~2023lli has a rare and long-duration bump in the UV/optical rising phase, which has the longest separation time from the main peak among TDEs. Moreover, weak and intermittent X-ray emission was detected 34 days after the main UV/optical peak. In this section, we investigate the origin of the bump and the X-ray emission mechanism. 

\subsection{The Origin of Optical Early Bump and Peak}

\subsubsection{Comparison with Bumps in Other TDEs}

\begin{figure}
    \centering
    \includegraphics[width=0.47\textwidth]{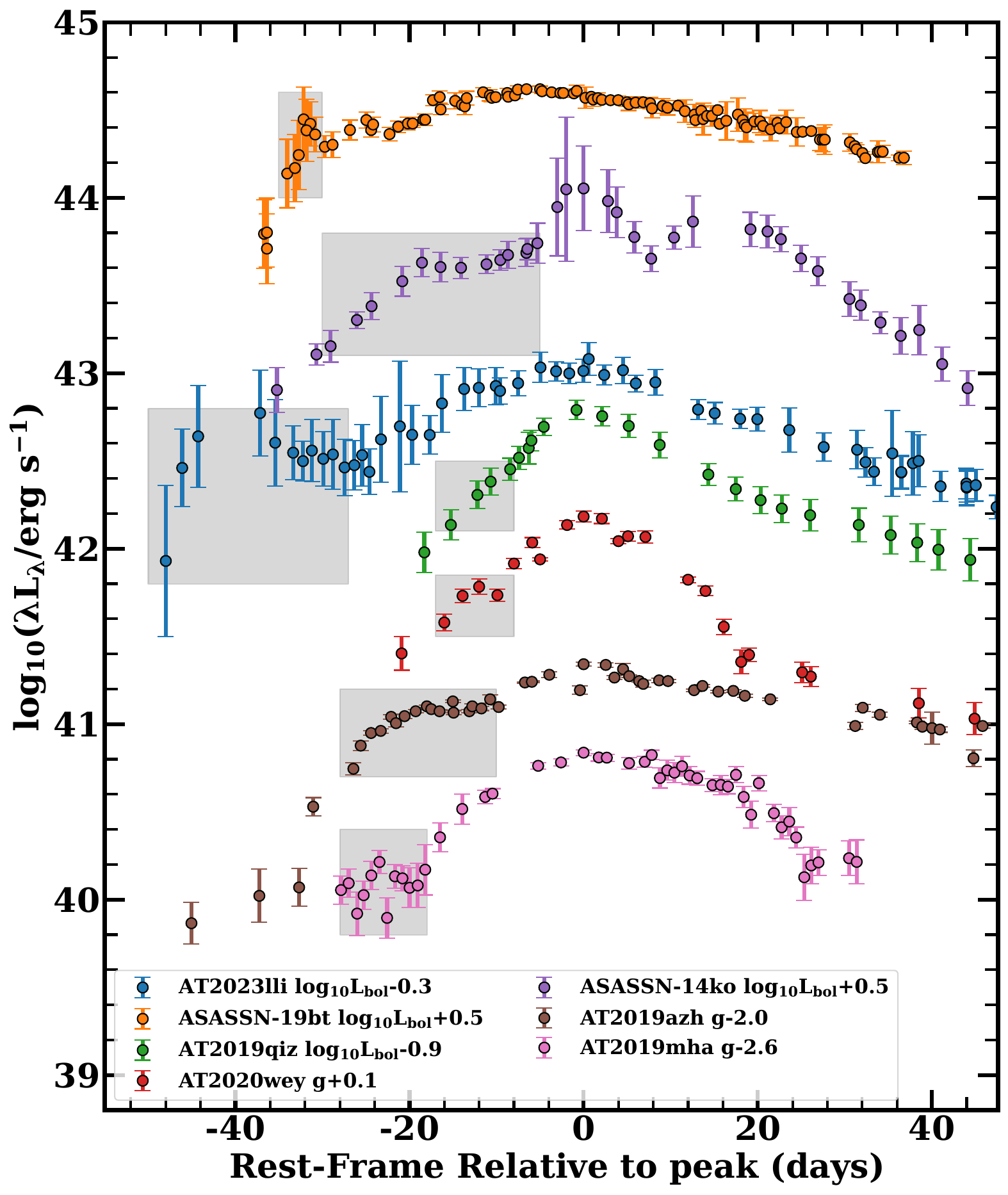}
    \caption{The comparison of TDEs with bump or precursors (marked by shadowed region). We derived the published data of AT~2020wey \citep{Charalampopoulos2023}, AT~2019azh \citep{Liu2022}, ASASSN-19bt \citep{Holoien2019}, AT~2019qiz \citep{Hung2021} and AT~2019mha \citep{Wangyb2023}. }
    \label{fig:TDE_bump}
\end{figure}

Early bumps in the UV/optical light curves are not rare in TDEs that have a well-covered rising phase light curve~\citep{Wangyb2023}. In Figure~\ref{fig:TDE_bump}, we compare the bump observed in AT~2023lli with previously reported cases for TDEs or TDE candidates. ASASSN-19bt, AT~2019azh, and AT~2020wey all display bumps that occur within 35 days of the main UV/optical peak~\citep{Holoien2019,Charalampopoulos2023,Faris2023}. \citet{Wangyb2023} reported early bumps that last around 10 days and occur within 30 days from the primary peak in ASASSN-18ap, AT~2019mha, and AT~2019qiz. The latter authors discussed several possibilities for the origin of the bumps, including the emission of unbound debris, the stream-stream collision, the vertical shock compression during the first passage, and the shock breakout of the debris collision. 
In contrast, the pTDE candidate ASASSN-14ko recently showed noticeable bumps before the main UV/optical peaks and rebrightenings after that. This can be interpreted as a consequence of the collision between the stream debris and the extended disk in the partial TDE scenario~\citep{Huang2023b}, which distinguishes it from AT~2023lli and other TDEs.

\subsubsection{Self-crossing of Stream Debris during Circularization and Delayed Accretion?}
Thanks to the prominent optical early bump, AT~2023lli has been extensively observed in multiwavelength observations including Swift/XRT, making it a unique object with the earliest X-ray observations among all optical TDEs, starting from 40 days (rest-frame) prior to its peak. It is worth noting that ASASSN-19bt, the only TDE known to possess X-ray observations as early as 1 month before the peak, also exhibits a tentative early bump in its bolometric light curve (\citealt{Holoien2019}, see also Figure~\ref{fig:TDE_bump}). Interestingly, neither TDE is X-ray detectable throughout the entire rising phase, yet AT~2023lli shows delayed X-ray emission in the late declining stage.

\begin{figure*}
    \centering
    \includegraphics[width=\textwidth]{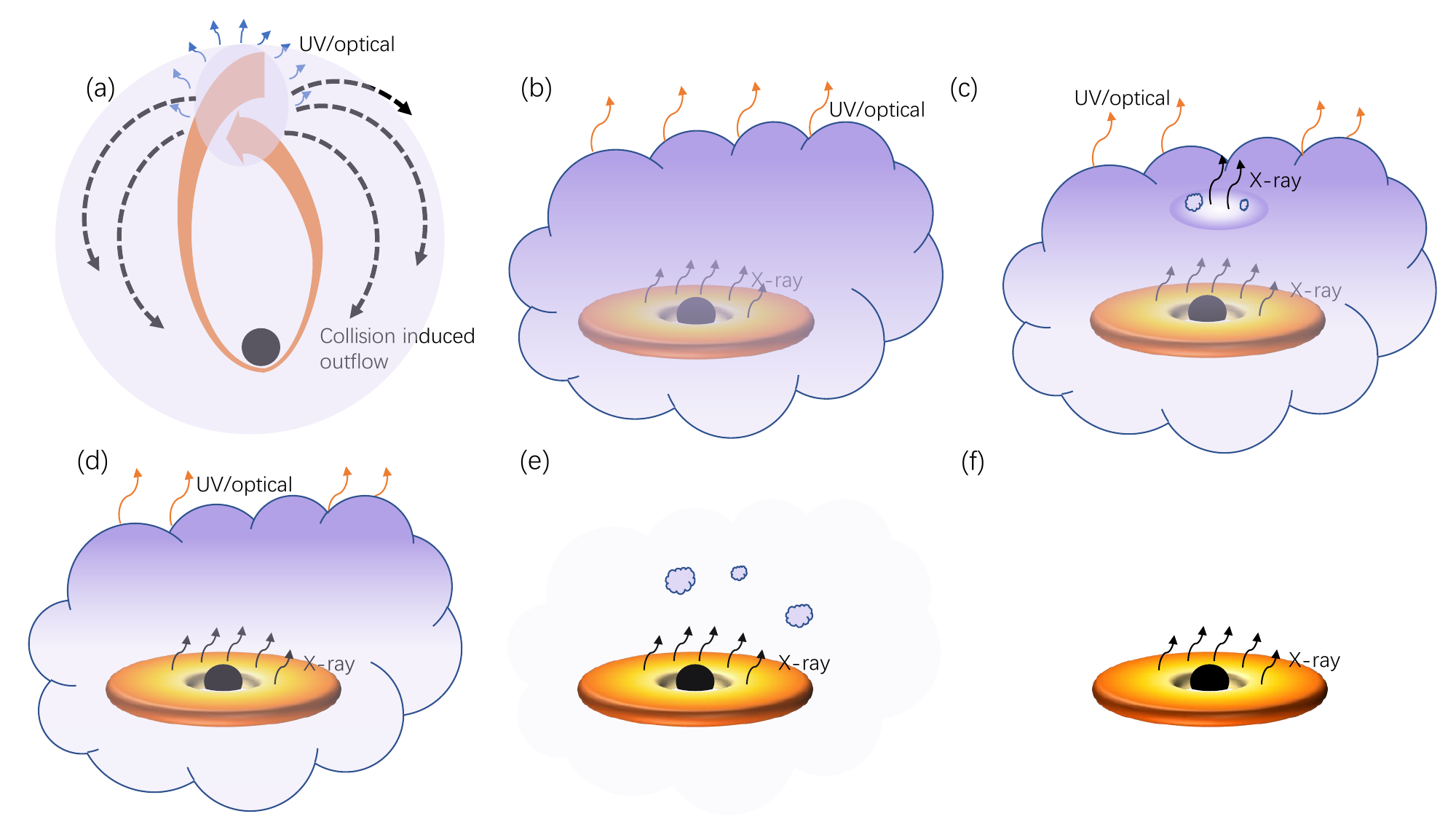}
    \caption{Schematic illustration of the processes that occurred in AT~2023lli, ordered in a chronological sequence from a to f. (a) A TDE occurred and the collision between streams produced UV/optical emission, which resulted in the UV/optical bump. This process also produced the outflow, with the material falling back to the vicinity of the black hole and wrapping around it. (b) As the accretion disk formed, it was obscured by surrounding materials, which absorbed the X-rays and produced the UV/optical emission through reprocessing, giving rise to the UV/optical main peak. (c) The materials expanded, and because of their inhomogeneous distribution, a locally optically thin region appeared, allowing X-ray photons to ``leak'' out. Some clumpy materials around caused episodic X-ray emission. (d) The optically thin region rotated and obscured the X-ray emission again. (e) The overall environment became optically thin, but some clumpy material still existed and obscured the X-rays, resulting in episodic emission. (f) The clumpy material disappeared, and X-rays could be observed continuously.  }
    \label{fig:model}
\end{figure*}

We propose a two-phase scenario to explain the optical and X-ray behaviors observed in this unique TDE. A chronological scheme is illustrated in Figure \ref{fig:model}.
A star is initially tidally disrupted into stream debris, which then undergoes a circularization process to form the accretion disk. Apsidal precession causes streams to self-cross, producing UV/optical emission~\citep{Piran2015}. This process may explain the early bump observed in AT~2023lli. A new disk forms from the stream debris as it loses energy in the stream-stream collision and circularizes, resulting in the main UV/optical outburst. However, the X-rays from the accretion disk are blocked by the reprocessing layer, probably in the form of collision-induced outflow in the early stage~\citep{Lu2020} and super-Eddington-accretion-induced outflow around peak~\citep{Metzger2016,Roth2016,Dai2018}. 

The UV/optical light curves of AT~2023lli resemble those of AT~2019avd, which also exhibit two peaks. They were interpreted as the result of stream circularization and delayed accretion, respectively \citep{Chen2022,Wang2023}.  Simulations suggest that when a TDE occurs, the disk wind will collide with the inner edge of the warped stream debris. The interaction produces a sharp peak in the rising phase of the TDE light curves \citep{Calderon2024}. This could be one of the possible origins of the bump in AT~2023lli.

\subsubsection{Double TDE?}
When the mass center of a binary passes too close to a black hole, it can become tidally separated, ultimately leading to the disruption of both stars. This phenomenon is known as a double TDE and may account for approximately 10\% of all TDEs~\citep{Mandel2015}. Simulations suggest that double TDEs may exhibit a dual-peak pattern in their light curves~\citep{Mandel2015}. with the significance of this feature depending on the mass ratio of the two disrupted stars~\citep{Mainetti2016}. Furthermore, in double TDEs, the separation between the two peaks is typically less than 150 days~\citep{Wu2018}. Therefore, the prominent bump and peak observed in the UV/optical light curve may also be attributed to a double TDE. During a double TDE, the production of two streams is possible, and there is a 44\% probability that these streams collide with each other~\citep{Bonnerot2019}. This collision could result in an outflow similar to the self-interaction of streams~\citep{Lu2020}, contributing to delayed X-ray emission through a reprocessing process. The bump and main outburst of the UV/optical luminosity can be fitted by ``Gaussian rise and power-law decay'' model, which has been applied to optically-selected TDEs. However, apart from the duration, there is no significant difference in the multi-wavelength light curves between the bump and the main outburst. As a result, there is currently insufficient evidence to fully support the hypothesis that the structure in the UV/optical light curves is caused by a double TDE, although it cannot be entirely ruled out.

\subsection{The Origin of the Episodic X-Ray Emission}
Until now, two main scenarios have been proposed for the delayed X-ray emission: delayed accretion after circularization~\citep{Gezari2017} and structural changes of the layer that reprocessed the X-rays~\citep{Thomsen2022,Guolo2023}. The first X-ray emission of AT~2023lli is detected on MJD 60198 albeit at the 2.5$\sigma$ level. After a month of dimming, the X-rays were detectable again. However, in subsequent observations, X-rays were not consistently detectable, even with sufficiently long exposure times. Here, we aim to explore the episodic X-ray emission in this source.

In the reprocessing scenario, X-rays are only detectable when the obscured materials become optically thin. As the obscuring material expands outward, it is likely that the material distribution becomes inhomogeneous, causing local regions to become optically thin earlier and ``leak'' X-ray photons. Rotation of the materials surrounding the disk could induce X-ray rebrightening when the optically thin region faces the observer again. The schematic of the model can be seen in Figure \ref{fig:model}.

Taking into account the Keplerian rotation of the optically thin region, its period is $T=2\pi \sqrt{\frac{a^3}{GM_{\rm BH}}}$, where $a$ is the semi-major axis and $G$ is the gravitational constant. Assuming the time interval of the two initial X-ray emissions as the period, we can estimate the location of the optically thin region as 
\begin{equation}
a=2.83\times 10^{14} \rm cm \left(\frac{T}{30~d}\right)^{2/3}\left(\frac{M_{BH}}{10^6M_{\sun}}\right)^{1/3}.
\end{equation}
This result is of the same order of magnitude as the blackbody radius derived from UV/optical SED fitting, indicating that the ``X-ray leak'' is located near the outer boundary of the obscured materials. The initial appearance of the X-rays lasted only 2 days, the second lasted 11 days, and the third lasted much longer. This leads us to speculate that if the ``X-ray leak'' is from the same region, the area is expanding, and we can expect to observe longer-lasting X-ray emission in the future. Subsequent observations confirm this conjecture. A similar scenario was previously used to explain the variability of RXJ1301.9+2747~\citep{Middleton2015}.

Using X-ray spectral fitting, we derived the hydrogen density ($1.41_{-1.23}^{+2.06}\times 10^{21}\,\text{cm}^{-2}$) of the photoelectric absorber, albeit with a large error bar. This may suggest the existence of obscured material around the inner disk. However, the material was optically thin at that time, so we could not obtain a better constraint on its parameter. Unfortunately, we did not have a high-quality spectrum during the early times when the X-ray emission was detected, which would have been the best opportunity to detect the absorber.

\cite{Jankovivc2023} demonstrated that the Lense-Thirring effect leads to an offset collision between streams, which causes the outflow to depart from spherical symmetry. Moreover, this effect reduces the covered solid angle of outflow and allows the X-ray photons to escape from the disk without being absorbed. It could be a potential source of the inhomogeneity that triggers episodic X-ray emission.

The delay and episodic X-ray emission suggest the production of a reprocessing layer during the TDE, similar to the case in AT~2019azh \citep{Liu2022}, which involves the self-crossing of stream debris and delayed accretion. Our high-cadence multiwavelength observations can provide support for the scenario that the X-ray delay to the UV/optical bands originates from the reprocessing process.

\section{Conclusions}\label{sec:conclusion}

The advanced modern time-domain surveys are not only discovering TDEs with an increasing speed but also unveiling the diversity of TDEs in their temporal evolution. AT~2023lli, a nearby TDE at redshift 0.036, is exactly such a new member that exhibits notable features in its optical and X-ray light curves. In this study, we conducted continuous multiwavelength observations of it. These observations revealed characteristic properties, including an almost constant blackbody temperature, very broad Balmer and possible He~II 4686\,\AA\ emission. However, it also exhibits some distinctive features that have captured our attention. The UV/optical light curves of AT~2023lli show a prominent bump that lasts nearly a month in the early rising phase. On the other hand, the time interval between the bump peak and the primary peak is nearly two months, making both timescales the longest among all TDEs with detected bumps, albeit only a handful so far. X-ray emission was not significantly detected until the late UV/optical decline stage ($\sim100$ days postpeak), despite two weak episodic emissions earlier. The XMM-Newton ToO observation of AT~2023lli, triggered by us upon persistent X-ray detection, revealed X-ray spectra with a possible intrinsic absorption component, albeit with large uncertainties. It is worth noting that AT~2023lli exhibits the power-law decay with an index of -4.10, making it one of the fastest-declining optically selected TDEs so far.

Taking all these observations into account, we suggest that the UV/optical bump may be caused by the self-crossing of the debris streams of the TDE, while the delayed accretion produces the UV/optical main outburst. The self-crossing or circularization of the debris streams produce outflows, which surround the accretion disk and cover it, leading to the absorption of soft X-ray photons from the inner disk and the detection of only the UV/optical emission. Due to the inhomogeneous distribution of the obscuring material, some regions were optically thin, allowing some X-ray photons to ``leak'' in the decline phase of the UV/optical. As the obscuring material evolved to be optically thin, the X-ray photons could be gradually detected. We have also discussed the possibility of double TDEs, in which the two stars in the binary are disrupted in sequence by the SMBH, which is also a possible scenario.

The early bump of TDEs is a newly recognized but likely very common characteristic that has not been previously discovered, primarily due to the lack of high-cadence deep photometric surveys, but this situation will soon change. The newly constructed WFST, which provides the late-time light curves of AT~2023lli in this work, will enable us to more precisely characterize the early rising light curves and their statistics for future TDEs located in the carefully designed deep high-cadence field of WFST~\citep{WFST}. Additionally, the joint study of WFST and the recently launched Einstein Probe~\citep{Yuan2018EP} may even have the opportunity to obtain a contemporaneous high-cadence X-ray light curve, which will definitely help characterize those complicated X-ray behaviors like episodic X-ray emission. Eventually, they might offer us new insights into the origin of optical emission of TDEs.  \\

\noindent 
We thank the anonymous referee for providing valuable comments, which helped to improve the manuscript. This work is supported by National Key Research and Development Program of China (2023YFA1608100), the National Natural Science Foundation of China (grants 12393814,12233008,12073025,12192221,12025303), the Strategic Priority Research Program of the Chinese Academy of Sciences (XDB0550200, XDB41000000, XDB0550300), Anhui Provincial Natural Science Foundation (2308085QA32), the China Manned Space Project (No. CMS-CSST-2021-A13) and the Fundamental Research Funds for Central Universities (WK3440000006). L.L. acknowledges the support of the National Natural Science Foundation of China (grant No. 12303050), the China Postdoctoral Science Foundation (2023M743397), and the Fundamental Research Funds for the Central Universities. The authors appreciate the support of the Cyrus Chun Ying Tang Foundations. The authors express their gratitude to Ping Cai for her help in creating the figure. We also appreciate the members of the WFST operation and maintenance team for their support. The Wide Field Survey Telescope (WFST) is a joint facility of the University of Science and Technology of China, Purple Mountain Observatory. This research uses data obtained through the Telescope Access Program (TAP), which has been funded by the TAP member institutes. Observations with the Hale Telescope at Palomar Observatory were obtained as part of an agreement between the National Astronomical Observatories, the Chinese Academy of Sciences, and the California Institute of Technology. The authors acknowledge the support of the Lijiang 2.4m telescope staff. Funding for the telescope has been provided by the Chinese Academy of Sciences and the People's Government of Yunnan Province”. The authors acknowledge the use of public data from the Swift data archive. The authors thank the Swift ToO team for accepting our proposal and executing the observations. Based on observations obtained with XMM-Newton, an ESA science mission with instruments and contributions directly funded by ESA Member States and NASA. This work makes use of observations from the Las Cumbres Observatory global telescope network. This work has made use of data from the Asteroid Terrestrial-impact Last Alert System (ATLAS) project. The Asteroid Terrestrial-impact Last Alert System (ATLAS) project is primarily funded to search for near earth asteroids through NASA grants NN12AR55G, 80NSSC18K0284, and 80NSSC18K1575; byproducts of the NEO search include images and catalogs from the survey area. This work was partially funded by Kepler/K2 grant J1944/80NSSC19K0112 and HST GO-15889, and STFC grants ST/T000198/1 and ST/S006109/1. The ATLAS science products have been made possible through the contributions of the University of Hawaii Institute for Astronomy, the Queen’s University Belfast, the Space Telescope Science Institute, the South African Astronomical Observatory, and The Millennium Institute of Astrophysics (MAS), Chile.

\vspace{5mm}
\facilities{Swift, XMM-Newton, Hale, LCOGT, WFST}

\software{astropy \citep{2013A&A...558A..33A}, HEASoft \citep{2014ascl.soft08004N}, Xspec \citep{1996ASPC..101...17A}, Matplotlib \citep{2007CSE.....9...90H}.
          }

\bibliography{AT2023lli}{}
\bibliographystyle{aasjournal}

\end{document}